\documentclass[12pt,preprint]{aastex}

\newcommand{\thirteenco}{${\rm ^{13}CO}$}
\newcommand{\twelveco}{${\rm ^{12}CO}$}
\newcommand{\eighteenco}{${\rm C^{18}O}$}

\newcommand{\tbol}{$T_{\rm bol}$}
\newcommand{\lbol}{$L_{\rm bol}$}
\newcommand{\lsun}{$L_{\odot}$}
\newcommand{\msun}{$M_{\odot}$}

\slugcomment{Accepted to ApJ: 10 Oct 2006}

\shorttitle{Spitzer Observations of B59}
\shortauthors{Brooke et al.}

\begin{document}

\title{The Spitzer c2d Survey of Nearby Dense Cores: IV. Revealing the 
Embedded Cluster in B59}

\author{Timothy Y. Brooke\altaffilmark{1}, 
        Tracy L. Huard\altaffilmark{2},
        Tyler L. Bourke\altaffilmark{2},     
        A. C. Adwin Boogert\altaffilmark{1,14}, 
        Lori E. Allen\altaffilmark{2}, 
        Geoffrey A. Blake\altaffilmark{3}, 
        Neal J. Evans, II\altaffilmark{4},
        Paul M. Harvey\altaffilmark{4}, 
        David W. Koerner\altaffilmark{5}, 
        Lee G. Mundy\altaffilmark{6}, 
        Philip C. Myers\altaffilmark{2}, 
        Deborah L. Padgett\altaffilmark{7}, 
        Anneila I. Sargent\altaffilmark{1}, 
        Karl R. Stapelfeldt\altaffilmark{8},
        Ewine F. van Dishoeck\altaffilmark{9},
        Nicholas Chapman\altaffilmark{6},
        Lucas Cieza\altaffilmark{4},
        Michael M. Dunham\altaffilmark{4},
        Shih-Ping Lai\altaffilmark{6,12,13},
        Alicia Porras\altaffilmark{2},
        William Spiesman\altaffilmark{4},
        Peter J. Teuben\altaffilmark{6},
        Chadwick H. Young\altaffilmark{10},
        Zahed Wahhaj\altaffilmark{5},
        Chang Won Lee\altaffilmark{11}}

\altaffiltext{1}{Astronomy Department, MC 105-24, California
Institute of Technology, Pasadena, CA 91125; tyb@astro.caltech.edu}
\altaffiltext{2}{Smithsonian Astrophysical Observatory, 60 Garden
Street, Cambridge, MA 02138}
\altaffiltext{3}{Department of Geological and Planetary Sciences, MC 150-21, 
California Institute of Technology, Pasadena, CA 91125}
\altaffiltext{4}{Department of Astronomy, University of Texas at
Austin, 1 University Station C1400, Austin, TX 78712-0259}
\altaffiltext{5}{Department of Physics and Astronomy, Northern Arizona
University, Box 6010, Flagstaff, AZ 86011-6010}
\altaffiltext{6}{Department of Astronomy, University of Maryland,
College Park, MD 20742}
\altaffiltext{7}{Spitzer Science Center, California Institute of
Technology, CA 91125}
\altaffiltext{8}{Jet Propulsion Laboratory, MS 183-900, California
Institute of Technology, 4800 Oak Grove Drive, Pasadena, CA 91109}
\altaffiltext{9}{Leiden Observatory, PO Box 9513, 2300 RA Leiden, the
Netherlands} 
\altaffiltext{10}{Department of Physical Sciences, Nicholls State 
University, PO Box 2022, Thibodaux, LA 70310} 
\altaffiltext{11}{Korea Astronomy and Space Science Institute, 61-1,
Hwaam-dong, Yuseong-gu, Daejeon 305-348, South Korea} 
\altaffiltext{12}{Institute of Astronomy and Department of Physics,
National Tsing Hua University, Hsinchu 30043, Taiwan} 
\altaffiltext{13}{Academia Sinica Institute of Astronomy and Astrophysics,
P.O. Box 23-141, Taipei 106, Taiwan}
\altaffiltext{14}{NOAO Gemini Science Center, Casilla 603, La Serena,
Chile, SA} 

\begin{abstract}

Infrared images of the dark cloud core B59 were obtained with the {\it
Spitzer Space Telescope} as part of the ``Cores to Disks'' Legacy
Science project.  Photometry from 3.6-70 $\mu$m indicates at least 20
candidate low-mass young stars near the core, more than doubling the
previously known population.  Out of this group, 13 are located within
$\sim0.1$ pc in projection of the molecular gas peak, where a new
embedded source is detected.  Spectral energy distributions span the
range from small excesses above photospheric levels to rising in the
mid-infrared.  One other embedded object, probably associated with the
millimeter source B59-MMS1, with a bolometric luminosity L$_{\rm bol}
\sim 2$ L$_\odot$, has extended structure at 3.6 and 4.5 $\mu$m,
possibly tracing the edges of an outflow cavity.  The measured
extinction through the central part of the core is A$_V$ $\gtrsim$ 45
mag.  The B59 core is producing young stars with a high efficiency.

\end{abstract}


\section{Introduction}~\label{sec:intro}

One component of the {\it Spitzer Space Telescope} ``Cores to Disks''
({\it c2d}) Legacy Science program (Evans et al. 2003), is imaging of
nearby dark cloud cores at infrared wavelengths ranging from 3.6 to 70
$\mu$m.  The program takes advantage of {\it Spitzer's} high
sensitivity (Werner et al. 2004) to survey young stars down to
$\sim0.001$ L$_\odot$ in these cores.  Five large molecular clouds are
also being studied by {\it c2d}.  This will allow comparisons to be
made between the young star populations in isolated cores and large
molecular clouds. {\it Spitzer's} high sensitivity in the mid-infrared
makes it ideally suited to studying the intermediate stages of
star-formation, particularly in dense cores, where dust extinction can
obscure the light at visible and near-infrared wavelengths.

This paper presents highlights of the {\it c2d} observational data of
the dark cloud core B59.  Though it is a known site of star-formation,
there have been relatively few published studies of this core to date
and spectral information is apparently lacking for nearly all of the
objects discussed here.  The {\it Spitzer} data show the core to be
the site of a small cluster of young low-mass stars at a range of
evolutionary stages.

\section{B59}~\label{sec:B59}

B59 (Barnard 1927; Lynds 1962) is an irregularly-shaped dark cloud
roughly 30\arcmin ~$\times$ 15\arcmin ~centered around
$\alpha$(J2000)= $17^{\rm h}$ $11.4^{\rm m}$ and $\delta$(J2000)=
-27\arcdeg ~26\arcmin ~(Herbig 2005).  It sits at the NW end (in
galactic coordinates) of the Pipe Nebula, a set of filamentary dark
clouds located close to our line of sight to the Galactic Center and
above the galactic plane ($l=357\arcdeg$, $b=7.2\arcdeg$ for B59).
The nebula also lies close in projection to the Ophiuchus molecular
cloud.

The Pipe Nebula was mapped in the $J = 1-0$ lines of \twelveco,
\thirteenco, and \eighteenco ~with the NANTEN telescope using a
2.7\arcmin ~beam (Onishi et al. 1999).  They identified 14 \eighteenco
~cores.  Molecular outflow emission was detected only towards a single
core (their Core 1), within B59.  This core has roughly twice the
column density of the other cores.  The outflow was not clearly
identifiable as bipolar, but may be due to two or more different
outflows.  We will refer to this as the B59 core, and sometimes
loosely B59, in this paper.

The distance to B59 is uncertain.  The radial velocity of the
molecular gas is consistent with a distance similar to the Ophiuchus
cloud (Onishi et al. 1999), which is approximately $d = 125\pm25$ pc
(de Geus et al. 1989).  The distance to the Pipe Nebula was recently
estimated by combining extinctions with parallaxes (Lombardi et
al. 2006).  They obtained $d = 130^{+13}_{-20}$ pc, where the
uncertainties are 1$\sigma$.  This distance is adopted in this paper.

B59 is the only confirmed star-forming region in the Pipe Nebula to
date.  In addition to the CO outflow(s), there are at least 5
H$\alpha$ emission-line stars in or near the core: LkH$\alpha$ 346 (NW
and SE), KW 2, B59-2, and LkH$\alpha$ 347 (Cohen \& Kuhi 1979; Herbig
\& Bell 1988; Kohoutek \& Wehmeyer 2003; Herbig 2005).  Two other
H$\alpha$ emitters, KK Oph and V359 Oph, are farther from the core
center, $\sim20$\arcmin ~and $\sim36$\arcmin, or 0.8 and 1.4 pc in
projection, but may also be associated with B59 (Herbig 2005).

There are 3 bright IRAS sources in the Point Source Catalog within
$\sim5.5$\arcmin ~of the center of the core with spectral energy
distributions consistent with young stellar objects, according to the
criteria of Lee \& Myers (1999): IRAS17079-2719 (associated with
LkH$\alpha$ 346), IRAS17081-2721, and IRAS17082-2724.  Because of the
proximity of the emission-line stars and IRAS sources, B59 is
categorized by the {\it c2d} team as a ``starred'' core.  A 1.3 mm
continuum survey revealed an additional probable protostellar object,
B59-MMS1 (Reipurth et al. 1996), not associated with any of the IRAS
sources.

There have been several studies of B59 that have focused on the
multiple stars.  B59-1 and the emission-line star, LkH$\alpha$ 346,
are probable triple systems (Chelli et al. 1995; Koresko 2002).
Another emission-line star, B59-2, is at least double (Reipurth \&
Zinnecker 1993; Chelli et al. 1995), and may be triple (Koresko 2002).

\section{Observations}~\label{sec:obs}

B59 was observed with the {\it Spitzer} Infrared Array Camera, IRAC
(Fazio et al. 2004), on UT 2004 September 3.9 and 4.2 under
Astronomical Observation Requests (AORs) 5131776 and 5132288.  The
pixel size is 1.2\arcsec.  Observations at two different epochs
allowed identification of asteroids.  The IRAC 3.6 $\mu$m and 5.8
$\mu$m arrays observe the same field simultaneously, and the 4.5
$\mu$m and 8.0 $\mu$m arrays a different field.  The observations at
each epoch consisted of a 3 $\times$ 4 map with small overlap regions.
For each map there were two dithers of 12-sec exposures, preceded by
short high-dynamic-range 0.6-sec exposures (HDR).  The HDR
observations allowed photometry of sources that saturated at longer
integration times.  The typical effective observation time was 48 sec.
The point-source flux limits for reliable detection were roughly 0.03,
0.03, 0.10, 0.15 mJy at 3.6, 4.5, 5.8, 8.0 $\mu$m, respectively.

Observations of B59 at 24 and 70 $\mu$m using the Multiband Infrared
Photometer for Spitzer, MIPS (Rieke et al. 2004), were obtained on UT
2005 April 6.4 and 7.0 (AORs 9409280 and 9438720).  The pixel size is
2.5\arcsec ~at 24 $\mu$m and 10\arcsec ~at 70 $\mu$m.  The
observations were 3 $\times$ 3 photometry maps with 1 cycle of 3 sec
observations at 24 $\mu$m, and 3 cycles of 3 sec at 70 $\mu$m.  The
second epoch was shifted by roughly 6\arcmin, to increase coverage at
70 $\mu$m.  Typical observation times were 96 sec at 24 $\mu$m and 100
sec at 70 $\mu$m.  The point-source flux limits for reliable detection
were roughly 0.5 mJy at 24 $\mu$m and 100 mJy at 70 $\mu$m.

The data presented here are from the {\it Spitzer Science Center's}
(SSC) pipeline version S11.4.0.  The data were further processed by
the {\it c2d} team to correct artifacts, where possible.  This was
followed by creation of new mosaics, source photometry up to 24
$\mu$m, bandmerging for the detected wavelengths, and preliminary
source typing with extinction estimates.  Counterparts in the 2MASS
catalog were assigned if found.  The procedures are described in Evans
et al. (2005).  Photometry at 70 $\mu$m was done separately with the
SSC's Mopex package, release version 030106 (Makavoz \& Marleau 2005).

One exception to these procedures is that three sources in the B59
field saturated at 24 $\mu$m (1, 7, and 11 in Tables 1 and 2) and one
source saturated at 70 $\mu$m (source 11).  For these sources,
photometry was done by fitting the radial profile in the wings with a
PSF obtained from an unsaturated bright point source in the field.
The flux uncertainty due to this fitting is estimated to be 20 \%.

The {\it Spitzer} photometry is summarized in Table 2.  The
uncertainties given do not include absolute calibration uncertainties.
These are currently estimated by the SSC to be 10 \% for IRAC, 10 \%
for MIPS 24 $\mu$m, and 20 \% for MIPS 70 $\mu$m.

The results in this paper are limited to the region where all IRAC
bands and the MIPS 24 $\mu$m areas overlap, about 170 arcmin$^2$.  The
MIPS 70 coverage is a subset of this, about 60 \% of the area.

\section{Results}~\label{sec:results}

An overview of the young star cluster in B59 is provided by the IRAC
3-color image and the MIPS 24 $\mu$m image in Fig. 1$a$ and
Fig. 1$b$.  Source 10 (Fig. 1$c$) is near the peaks in the
molecular gas and in the dust extinction, which are discussed below.

Selection of candidate young stars in B59 was based on infrared
excesses relative to photospheric colors, after correction for
extinction.  The excess method is needed because of the absence of any
systematic study of the core membership.  Correction for extinction is
vital due to the high dust column densities toward the core.

As part of its bandmerging, {\it c2d} checks whether a reddened
stellar photosphere can fit the 2MASS and {\it Spitzer} fluxes at all
wavelengths available for each source, within the uncertainties.  When
this was successful, the source was dropped from consideration as a
candidate young star here.  Note that some of these may in fact be
young stars, though with no detected infrared excesses.

For the remaining sources, which could be young stars, main-sequence
or post-main-sequence stars with excesses, or galaxies, we imposed a
24 $\mu$m flux lower limit, F$_{\nu}$(24 $\mu$m) $>$ 3 mJy.  This
tends to cut out galaxies as the number of galaxies brighter than this
is $\sim50$ deg$^{-2}$ (Marleau et al. 2004; Papovich et al. 2004), or
$\sim2$ in our field.  (Two of our candidates have no measured 24
$\mu$m fluxes due to confusion with nearby sources, but their expected
24 $\mu$m fluxes would easily make this cut.)

The galactic background counts toward B59 are less certain.  The model
of Wainscoat et al. (1992) indicates that for F$_{\nu}$(24 $\mu$m) $>$
3 mJy, the number of non-photospheric galactic sources, primarily
O-rich and C-rich asymptotic giant branch (AGB) stars, should be
small, $\sim13$ deg$^{-2}$, or $\sim0.6$ in our field.  But models can
serve only as a guide for small areas.

Lombardi et al. (2006) identified a population of AGB stars behind the
Pipe Nebula from their location in the 2MASS color-color diagram.
These are bright objects, peaking at K $\sim7$, most likely in the
galactic bulge.  If, as suggested by Lombardi et al., they are similar
to the OH/IR star sample of Jim\'enez-Esteban et al. (2005), then they
would be easily detectable at 24 $\mu$m, F$_{\nu}$(24 $\mu$m)
$\gtrsim$ 1 Jy.  But they are too few in number to significantly
contaminate the young star candidate sample; at $b = 7.2\arcdeg$,
there are $\sim3$ deg$^{-2}$, or $\sim0.14$ in our field.

To be conservative, we also cut out sources with excesses which
appeared close to photospheric levels.  We required at least one flux
to be a factor $\sim4$ higher than the expected photosphere at that
wavelength.  This eliminated 4 more objects.  In doing so, we may have
dropped some young stars with weak dust excesses.  The motivation is
that {\it c2d} is finding that there are some sources with small IR
excesses toward the Serpens cloud that are background AGB stars.

The 20 candidate young stars are listed in Tables 1 and 2.  Positions
are shown in Fig. 1$c$.  Note that 13 of the objects lie within
$\sim0.1$ pc in projection of the molecular gas peak (see Fig. 1).  We
cannot completely eliminate the possibility that some of the proposed
young stars are background objects without further data.  So we refer
to the set as candidate young stars.  But some have H$\alpha$ in
emission (Section 2), and the majority are clustered together close to
the center of the core.

For the candidate young stars, we re-evaluated their extinction taking
into account the possibility of excess emission in the near-IR.  There
is no systematic visible photometry of the cloud members, nor is there
information on spectral types or effective temperatures for most of
the objects.  We rely on typical dereddened near-IR colors of young
stars.

We assumed that the young stars with all 3 2MASS fluxes in our sample
have intrinsic JHK$_s$ colors that fall on the locus of the sample of
T Tauri stars in Taurus, whose dereddened colors were estimated by
Meyer et al. (1997), see Fig. 2.  This requires that we are sampling
similar spectral types and dust properties as in Taurus.  It is
difficult to test this without spectroscopic data on the stars.  But
the assumption has been used for other young clusters, e.g. IC 348
(Muench et al. 2003) and NGC2024 (Haisch et al. 2001), in the absence
of better constraints.  And Luhman (2004) shows consistency of
extinction estimates done this way compared to those obtained from
visible spectra for young stars in Cham I.

We adopted the dust extinction law of Indebetouw et al. (2005), which
extends from 1.25-8.0 $\mu$m.  The 1.25-2.2 $\mu$m extinction of the
Pipe Nebula agrees with that of Indebetouw et al. within the
uncertainties (Lombardi et al. 2006).  Using this dust extinction law,
we allowed the young stars to deredden to the locus of JHK$_s$ colors
of the T Tauri stars, or to the nearest extrapolated point (Fig. 2).
Uncertainties from both the data and the locus envelope are propagated
to the extinction estimate (Table 1).  By convention, the extinction
is given by A$_V$ in magnitudes, whereas we do not know the extinction
law in this region down to visible wavelengths.  We only estimate
near-infrared values of the extinction, e.g. A$_K$.  We converted
using A$_V$/A$_K$ = 8.9 from the Rieke \& Lebofsky (1985) extinction
law, but note that our actual estimates are A$_K$.  The adopted
Indebetouw et al. extinction law appears appropriate for B59 in the
JHK$_s$ bands, as the bright stars in the field follow it (Fig. 2).

For source 8, with only HK$_s$ data, we deredden to the center of the
T Tauri locus, and assign a larger uncertainty.  The embedded objects
(10, 11) have insufficient near-IR data to correct for extinction by
this method.

The dereddened spectral energy distributions (SEDs) are shown in
Fig. 3.  The 24 $\mu$m extinction, A$_{24}$/A$_K$ = 0.24, was taken
from the Weingartner \& Draine (2001) model which fits the Indebetouw
et al. points (R$_V$=5.5, Case B). 

The 2-10 $\mu$m or 2-20 $\mu$m slope is commonly used to classify
young stars in the IR.  Following earlier authors (Lada 1987; Wilking
et al. 2001; Greene et al. 1994) we define the spectral index as
$\alpha=d{\rm log}(\lambda F_\lambda)/d{\rm log}\lambda$.  We define
the IR spectral classes as follows: $\alpha \geq 0.3$ is a Class I
source, $0.3 > \alpha \geq -0.3$ a ``Flat'' type, and $\alpha < -0.3$
a Class II.  These agree with Greene et al. (1994), except that the
Class III designation is restricted to sources with practically
photospheric slopes after dereddening, $\alpha \sim -3$ (Kenyon \&
Hartmann 1995).  Because of our selection criteria, we have no Class
III sources in the present sample.  We use the 2MASS K$_s$ and IRAC 8
$\mu$m fluxes as endpoints, unless noted in Table 2.  Slopes and IR
class are given in Table 2 for the observed fluxes uncorrected for
extinction and with correction for extinction.  (Source 1 after
correction is on the border of Flat and Class II, but the complete SED
indicates it is most like the Flat objects.)

The similarity of the dereddened SEDs to known young stars in other
star-forming regions gives us confidence that the A$_V$ values are
reasonable and that all of the objects are young stars.  Note in
particular (Fig. 3) the strong similarity of most of the Class II
sources with $\alpha \geq -1.6$ to the median Taurus Class II SED from
the sample defined by D'Alessio et al. (1999).

\subsection{Extinction Map}~\label{sec:ext}

There are sufficient numbers of background stars detected through the
B59 region to provide an approximate extinction map.  The H-K color
excess was adopted as a measure of extinction (Lada et al. 1994).
This is an effective technique as the spread in H-K colors of
main-sequence stars is small ($\sim0.3$ mag).  We require 2MASS colors
at present, so some highly extincted background stars (for which we
could in principle estimate A$_V$ from {\it Spitzer} colors alone)
could not be used.  But there are enough 2MASS sources for a map, and
the accuracy and wide use of this method led us to adopt it.

We assumed a mean intrinsic color of H-K = 0.175 for the background
stars, which is the mean value for stars in the Pipe Nebula control
field of Lombardi et al. (2006), obtained from M. Lombardi
(priv. comm.).  The conversion to A$_V$ as described above is A$_V$ =
16.8E(H-K), where E(H-K) is the color excess.  The factor 16.8 differs
from the commonly used 15.9, due to the adoption of the Indebetouw et
al. (2005) near-IR extinction law.  Note that the maximum uncertainty
in intrinsic color translates into an uncertainty in A$_V$ for an
individual star of 2.5 mag, which is small enough to not affect our
results.

The spatial resolution of the extinction map was set by the need for
several stars in each spatial bin.  This required boxes of size
$\sim100$\arcsec, and the map was oversampled by a factor 2.  The
median A$_V$ in each bin was taken to avoid foreground objects and the
scatter gave the uncertainty.  The extinction map is shown in
Fig. 1$d$, and interpolated values at the positions of the candidate
young stars in Table 1.

The agreement between the total extinction A$_V$ estimates and those
towards individual young stars is not particularly good.  Most of the
extinction map A$_V$ values are higher than those derived by
dereddening individual objects.  This is reasonable since many of the
objects could be near the front of the cloud.  Two are lower, which
may be explained by localized dust.

The measured peak extinction from the background stars is A$_V$ $\sim$
45 magnitudes, though the true value is likely higher due to the
coarse spatial resolution.  We estimated the molecular hydrogen column
density at the peak using N(H$_2$) = $9.4 \times 10^{20}$ A$_V$
(Kandori et al. 2005), giving N(H$_2$) = $4.2 \times 10^{22}$
cm$^{-2}$ with a 50 \% uncertainty.  For the \eighteenco ~core in B59,
Onishi et al. (1999) estimated N(H$_2$) = $1.5 \times 10^{22}$
cm$^{-2}$ from the strength of the \eighteenco ~line.

\subsection{Comparison to Gas}~\label{sec:gas}

In addition to the CO data of Onishi et al. (1999), data for B59 in CS
(J=2--1) and N$_2$H$^+$ (J=1--0) have been obtained at FCRAO with an
effective beam size of 60\arcsec ~(C. DeVries, in preparation).  CS
integrated intensity contours are shown in Fig. 1$e$.  Source 10 lies
close to the CS molecular gas peak.  The N$_2$H$^+$ emission (not
shown) is centered near this area, but appears extended NE-SW,
i.e. similar to the extinction.  No strong, well-organized outflow or
infall signatures were detected for either species in these data.

The positions of the CO outflows within our mapped region from Onishi
et al. (1999) are marked in Fig. 1$e$.  These are the positions of the
peak contours in the red and blue channels defined by those authors.
However the large beam size of these observations (2.7\arcmin) makes
detailed comparison to the {\it Spitzer} sources difficult.  It
appears that some of the CO outflows are aligned with source 11,
rather than source 10 which is closer to the extinction peak (see
Sec. 4.4).

The total gas mass in the B59 core estimated from the \eighteenco
~line is $\sim20$ \msun ~within a radius of $\sim0.15$ pc (Onishi et
al. 1999, both adjusted for the different assumed distances).

The total gas mass in the core can also be estimated from the dust
extinction of background stars under the assumption that dust within
the core provides most of the extinction.  The conversion from A$_V$
to N(H$_2$) for typical interstellar dust is given above.  For a
standard cloud composition with mean atomic weight per H atom, $\mu =
1.37$ (Lombardi et al. 2006), the total core mass integrated within
the confines of the \eighteenco ~core is $M_{\rm core} = 25 \pm 1$
\msun, where the uncertainty does not include the distance
uncertainty.

These two estimates of core gas mass are in good agreement,
considering the different spatial resolutions.  Some freeze-out of CO
is probable in the dense regions.  Although solid CO has not yet been
observed toward B59, sources near the center of the core have deep
solid H$_2$O and CO$_2$ bands (A. Boogert, in prep.).

\subsection{Class 0/I and I Sources and Bolometric Luminosities}~\label
{sec:mms1}

Based on their SED's there are two deeply embedded objects near the
center of the B59 core.  Source 11 = 2MASSJ17112317-2724315 (hereafter
2MASS171123) lies $\sim$15\arcsec ~E of the quoted position of the 1.3
mm source B59-MMS1 observed with a 22\arcsec ~HPBW beam (Reipurth et
al., 1996) and the two are probably the same.  The total 1.3mm flux
integrated to the 8$\sigma$ contour was 725 mJy.  It was recently
detected at 350 $\mu$m with the SHARCII bolometer array at the Caltech
Submilllimeter Observatory (Wu et al. 2006, submitted to ApJ).  The
flux in a 40\arcsec ~diameter aperture is $45.2 \pm 8.0$ Jy.

With most of the SED for this source in hand (Fig. 4), we estimated
the bolometric luminosity, \lbol $=2.2 \pm 0.3$ \lsun, the bolometric
temperature, \tbol$=70 \pm 10$ K (Myers \& Ladd 1993), and the ratio
of sub-mm ($\gtrsim350$ $\mu$m) to bolometric luminosity ($L_{\rm
smm}/$\lbol $=0.03 \pm 0.01$).  The uncertainty in luminosity does not
include the distance uncertainty.

In addition to the Class I-III spectral classes for young stars
discussed above, a fourth class, Class 0, is used to define cold
sources.  Two common criteria for a Class 0 source are that \tbol$<70$
K and $L_{\rm smm}/$\lbol $ > 0.5$ \% (Andr\'e et al. 2000).

2MASS171123 lies somewhere near the border of this definition, so may
be described as a Class 0/I source.

As noted by Young et al. (2004) and others, heating of dust by the
interstellar radiation field may contribute to the fluxes of
low-luminosity protostars longward of $\sim100$ $\mu$m.  Thus high
values of $L_{\rm smm}/$\lbol ~may be common for low-luminosity
protostars at some point during their lifetime, and some revision of
this criterion may be needed.

The other nominal Class I object, source 10, is close to the gas and
dust extinction peaks.  It lacks a complete SED at present, but
appears spectrally similar to source 11 in the mid-IR, so we estimate
a bolometric luminosity by scaling, \lbol $\sim0.6$ \lsun, but this is
only a crude estimate.  We will leave its designation as Class I for
the present time, pending mm and sub-mm observations.

We are not able to determine directly the final stellar masses for
these two Class 0/I or I souces.  The models of Myers et al. (1998)
suggest that source 11 and source 10 will evolve to form stars with
0.5 and 0.3 \msun, and have ages of $t\sim0.1-0.2$ Myr.

It is probable that one or both of the embedded sources near the
center of the core we detect with {\it Spitzer} are the youngest
objects and drive the CO outflows seen by Onishi et al. (1999).
Further study of these objects and their possible disks and outflows
with IR and mm and sub-mm interferometry is needed.

Bolometric luminosities (Table 2) for the rest of the candidate young
stars were estimated as follows.  For the Class II sources with
$\alpha < -1.6$, late-type photospheres were fitted to the dereddened
J-band fluxes.  The IR excesses add only marginally to the bolometric
luminosity.  For the Class II sources with $\alpha \geq -1.6$, we used
the absolute J band flux vs. bolometric luminosity relation from
Greene et al. (1994).  For the Flat sources, we have no good estimate;
only lower limits were obtained by integration under the observed
fluxes.

Estimated bolometric luminosities for the Class II sources range from
0.1-1.8 \lsun.  The central stars provide the bulk of the contribution
to the bolometric luminosity for these objects.  Stellar evolutionary
models (e.g. D'Antona \& Mazzitelli 1994; Siess et al. 2000) show that
for plausible ages of $t=0.5-1.0$ Myr, the stellar masses are
$\sim0.1-0.9$ \msun, corresponding to mid G to late M dwarfs.  There
are no clear brown dwarf candidates in the present sample.

\subsection{Extended Structure around 2MASS171123}~\label{sec:disk} 

The Class 0/I source 2MASS171123 has extended structure at 3.6 and 4.5
$\mu$m (Fig. 5), extending out $\gtrsim30$\arcsec ~(0.02 pc) to the NE
and SE.  There is a trace of extension at 5.8 $\mu$m also.  These arcs
may trace the edges of an outflow cavity in reflected light.  The
morphology suggests a disk inclined to the line of sight.  A possible
third extension to the W is confused with a superposed background
source.

The apparent outflow axis defined by the E lobes is $\sim50\arcdeg$ ~E
of N.  This lines up with some of the CO outflow peaks (Onishi et
al. 1999; see Fig. 1$e$).  But the low spatial resolution of the CO
data makes a firm identification of specific outflows with this
source difficult.

The arcs are unlikely to be due to H$_2$ emission, because the {\it
Spitzer} IRS spectrum of the object shows no evidence of H$_2$
emission lines at longer wavelengths (A. Boogert, in preparation).

\subsection{Extended Emission}~\label{sec:glo}

The Spitzer images of B59 show two kinds of extended emission above the
background zodiacal emission.

First is the red (8.0 $\mu$m) emission seen most clearly in the NW
area of Fig. 1$f$, roughly 0.8 MJy/sr above background.  This is due to
galactic cirrus dust emission.  The same structure is seen in the IRAC
5.8 $\mu$m band near the detection limit (not shown).  The band ratio
is I(5.8)/I(8) $\approx$ 0.25.  The emission at 24 $\mu$m also broadly
traces the same area, with I(24)/I(8) $\approx$ 0.4.  These flux
ratios are in the range expected for galactic cirrus (e.g. Reach et
al. 2004), with the 8 $\mu$m flux due mostly to the emission in
the strong aromatic hydrocarbon bands at 7.7 and 8.6 $\mu$m.

The second extended structure is more interesting: a wide greenish
glow centered on the core, due to emission in the IRAC 4.5 $\mu$m band
with a contribution from the blue 3.6 $\mu$m band.  This emission is
just visible in Fig. 1$f$ (it may be easier to see in the online
version).  The band ratio is roughly I(3.6)/I(4.5) = 0.75.  This
emission is not detected at longer wavelengths.  The coincidence of
the emission with the cloud core is an indication that this is due to
scattered galactic light, sometimes called cloudshine (Foster \&
Goodman 2006).  

Cloudshine has been seen in many dense cloud cores at visible (Witt \&
Stephens 1974) and near-IR wavelengths (Lehtinen \& Mattila 1996;
Nakajima et al. 2003; Foster \& Goodman 2006).  The JHK data of Foster
\& Goodman (2006) have been used to study dust structure on
sub-arcsecond scales (Padoan et al. 2006).  It is reasonable to expect
cloudshine at longer wavelengths as well, though typical interstellar
grains will have decreased scattering efficiency at longer
wavelengths.  Longer wavelengths will in principle penetrate deeper
into the cloud, but the visibility will depend on the sensitivity of
the observations and the strength of the ambient galactic light.

This explanation is favored over H$_2$ line emission because of the
lack of an obvious jet or shock morphology, and because no H$_2$ lines
are seen in the IRS spectrum of source 10 (A. Boogert, in
preparation).

\section{Discussion}~\label{sec:disc}

The {\it Spitzer} data reveal the dark cloud core B59 to be a region
of active low-mass star formation containing a small cluster of young
stars at different evolutionary stages.  Some authors (e.g. Lada \&
Lada 2003) prefer to reserve the word cluster for groups of $\geq35$
objects, but we use the term in a general sense here.  Some notable
aspects and limitations of the {\it Spitzer} observations are
discussed in this section.

{\it Completeness for Weak Excesses--} The sensitivity of this
analysis to young stars with weak dust excesses is limited by the 24
$\mu$m flux cutoff of 3 mJy.  A hypothetical young star at A$_V$ = 10
with a factor 4 excess at 24 $\mu$m would not make the cut if it's
unextincted photosphere had a flux below 1 mJy.  For a main-sequence
object at $d=130$ pc, this corresponds to a mid-K dwarf.  For a more
likely age of $t\sim1$ Myr, this corresponds to an object that will
become roughly a mid-M dwarf (Baraffe et al. 2002).  Later spectral
types make the cut if they have more substantial excesses.  So the
sample defined here may be incomplete in Class II for late spectral
types with weak excesses.

{\it IRAC Color-Color Plot--} Many {\it Spitzer} studies of young
stars are possible with IRAC alone, e.g. Hartmann et al. (2005).
Fig. 6 shows an IRAC color-color plot for the candidate young stars in
B59 for comparison to IRAC studies of other star-forming regions.  The
colors plotted are as observed, though the IR spectral classes
indicated are after correction for extinction, where possible.  Table
2 has the IR spectral classes before correction for extinction.  The
adopted magnitude zero points were 278 Jy (3.6 $\mu$m), 180 Jy (4.5
$\mu$m), 117 Jy (5.8 $\mu$m), and 63.1 Jy (8.0 $\mu$m).

{\it Disk Properties and Age--} We have as yet no direct measure of age
(e.g. Li absorption lines) for any of the candidate young stars.  The
IR spectral classes give a rough measure of age, with Class I being
the youngest and Class III the oldest (Kenyon \& Hartmann 1995).

Many authors use a two-class definition for the IR spectral classes:
$\alpha>0$ for Class I, $\alpha<0$ for Class II.  For comparison in
the next paragraph, B59 has a two-class Class I to Class II ratio of
3:17, or 18 \%.

The two-class Class I/Class II ratio of 18 \% can be compared to {\it
Spitzer} results on other young clusters: 17-34 \% for the Tr 37
globule ($t\sim1$ Myr, Sicilia-Aguilar et al. (2006); Reach et
al. (2004)); 25 \% for NGC 7129 ($t\sim1$ Myr, Muzerolle et
al. (2004)); and 33 \% for the Spokes cluster in NGC 2264 ($t\sim0.5$
Myr, Teixeira et al. (2006)).  For two clusters in Perseus, Jorgensen
et al. (2006) find: 29 \% for NGC1333 ($t\lesssim1$ My) and 15 \% for
the older IC348 ($t\sim2$ Myr).

The above comparisons should be taken only as suggestive due to
several factors: some studies do no extinction corrections; there are
different sensitivities for different young star selection techniques;
and sensitivities differ because of different cloud distances.

We can also use an argument based on the predicted lifetimes of
embedded and T Tauri phases (Myers et al. 1987; Kenyon et al. 1990) to
estimate the onset of star-formation in B59.  For a fixed
star-formation rate of objects which evolve from Class I
(t$_{life}$ $\sim$ 10$^5$ yr) to Class II (t$_{life}$ $\sim$ 10$^6$
yr) the relative numbers suggest an age of $\sim$ 0.7 $\times$ 10$^6$
yr.  Clearly the number of objects in each bin in B59 makes this
argument only approximate.  And we have neglected potentially
important factors, e.g. the effect of the stellar environment on disk
lifetimes.
 
Multiplicity also complicates the interpretation.  The suspected young
star pair resolved with {\it Spitzer}, LkH$\alpha$ 346 NW and SE
(itself a probable double), have very different SEDs (our IR spectral
classes are Flat and Class II, respectively).  If coeval, the simple
model sketched above is wrong.  In addition, the frequency of disks
may be a function of spectral type, as indicated by a recent {\it
Spitzer} study of IC348 (Lada et al. 2006).

In Fig. 3, we compared the SEDs of the B59 Class II objects with
$\alpha \geq -1.6$ to a median of a sample of mostly Class II objects
in Taurus (D'Alessio et al. 1999).  The Taurus SED can be reasonably
fitted by optically thick disks with flaring at large radii (D'Alessio
et al. 1999).  With time, disks are expected to clear from inside-out
due to dust coagulation into planetesimals and dynamical clearing.
Evidence for such transition disks has recently been shown for IC 348
($t\sim2$ Myr, Lada et al.(2006)) and Tr 37 ($t\sim4$ Myr,
Sicilia-Aguilar et al. (2006)).  The fact that most of the SEDs of the
Class II sources with $\alpha \geq -1.6$ in B59 are similar to the
Taurus SED suggests a comparable age for the cloud, $t\sim0.7$ Myr
(Kenyon \& Hartmann 1995).

We see no clear ``transition objects'', such as CoKu Tau/4 (Hartmann
et al. 2005), those which have very little excess in the IRAC bands,
then a large 24 $\mu$m excess.

These comparisons taken together suggest an age for B59 of between 0.5
and 1 Myr.  Deep optical and near-IR photometric and spectroscopic
surveys of the cloud are needed to ensure the membership and measure
more accurate age indicators.

{\it Star Formation in B59--} To date, B59 is the only known
star-forming cloud in the Pipe Nebula.  Onishi et al. (1999) suggest
that this is due to its proximity to the B0 star $\tau$ Oph, with a
stellar wind having triggered star formation.

It is notable that the star formation efficiency of B59 is relatively
high.  Taking just the objects near the core, there are 13 young stars
in $\sim25$ \msun ~of dense gas, or 0.52 \msun$^{-1}$.  Of the 179
\eighteenco ~cores in nearby star-forming regions covered in the
NANTEN survey, only a few have a comparable efficiency (Tachihara et
al. 2002, and references therein).  By way of contrast, the Taurus
cloud as a whole has roughly 200 young stars in $\sim10^4$ \msun ~of
gas (Kenyon \& Hartmann 1995; Lada et al. 1993).

Assuming an average mass for the young stars in B59 of 0.5 \msun, the
star formation efficiency as defined by Lada \& Lada (2003),
M$_{star}$/(M$_{star}$+M$_{gas}$) = 0.21, close to the maximum of
$\sim0.3$ seen in their survey of young stellar clusters.

The surface density of young stars in the inner core region of B59 is
$\sim200$ pc$^{-2}$.  This is a lower limit to the young star
population as we require infrared excess for inclusion.  The surface
density is greater than those of 90 \% of the young clusters tabulated
in Lada \& Lada (2003), though not as high as much more massive
clusters such as S106 and the Trapezium.  However, the B59 core is
smaller than most of the Lada \& Lada sample.  Comparison to the
NANTEN core sample cited above again reveals only a few with
comparable young-star surface density.  Recent {\it c2d} observations
of another small cluster, L1251B, give a comparable stellar density to
B59 (Lee et al. 2006).

Compared to all of the cores in the NANTEN CO survey (Tachihara et
al. 2002), B59 is notable for its high gas column density, higher than
nearly all of the starless and single-star-forming cores, but similar
to their cluster-forming cores ($\gtrsim10$ stars), consistent with
our survey findings of a small cluster in formation.

As noted by Lada \& Lada (2003), our inventory of young, embedded star
clusters, even within a few hundred parsecs, is still incomplete.  Data
from the {\it Spitzer} Space Telescope are helping to complete that
inventory.

\section{Summary}~\label{sec:disc}

\noindent
1) At least 20 candidate young stars in the B59 core have been
identified from infrared excesses in {\it Spitzer} 3.6-70 $\mu$m
observations.  Spectral energy distributions including corrections for
extinction are presented.  They range from small excesses above
photospheric levels to rising in the mid-infrared.  The SEDs of most
of the Class II spectral types with spectral index $\alpha \geq -1.6$
are similar to the SED of the Class II sample in the Taurus cloud of
D'Alessio et al. (1999).

\noindent
2) Thirteen of the young star candidates are located within $\sim0.1$
pc in projection of the molecular gas and dust extinction peak.  Two
Class 0/I or I sources are identified.  One, 2MASSJ17112317, probably
associated with the millimeter source B59-MMS1, with a bolometric
luminosity L$_{\rm bol} \sim 2$ L$_\odot$, has extended structure at
3.6 and 4.5 $\mu$m, possibly tracing the edges of an outflow cavity.
The other is near the molecular gas and extinction peak and previously
unrecognized.

\noindent
3) The measured extinction through the central part of the B59 core is
A$_V$ $\gtrsim$ 45 mag.  There are extended infrared emission regions
due to galactic cirrus emission and scattered galactic light.

\noindent
4) Although further study of the young star population in B59 is
needed, the data to date suggest the B59 core is producing young stars
with high efficiency.

\acknowledgments

We thank M. Lombardi (ESO) for assistance, C. DeVries (CSU Stanislaus)
for making molecular line data available prior to publication, and
R. Gutermuth (CfA) for display scripts.  We thank the referee whose
comments inproved the paper.  This work is based on observations made
with the {\it Spitzer Space Telescope}, which is operated by the Jet
Propulsion Laboratory, California Institute of Technology, under NASA
contract 1407.  Support for this work, part of the Spitzer Space
Telescope Legacy Science Program, was provided by NASA through
contract numbers 1224608, 1230779, and 1230780 issued by the Jet
Propulsion Laboratory, California Institute of Technology, under NASA
contract 1407.  Astrochemistry in Leiden is supported by a NWO Spinoza
grant and a NOVA grant.  CWL acknowledges suport from KOSEFF grant
R01-2003-000-10513-0.  This publication makes use of data products
from the Two Micron All Sky Survey, which is a joint project of the
University of Massachusetts and the Infrared Processing and Analysis
Center/California Institute of Technology, funded by NASA and NSF.
This research has made use of the SIMBAD database, operated at CDS,
Strasbourg, France.

\clearpage

\begin{figure}[!t]
\includegraphics[angle=0, scale=0.8]{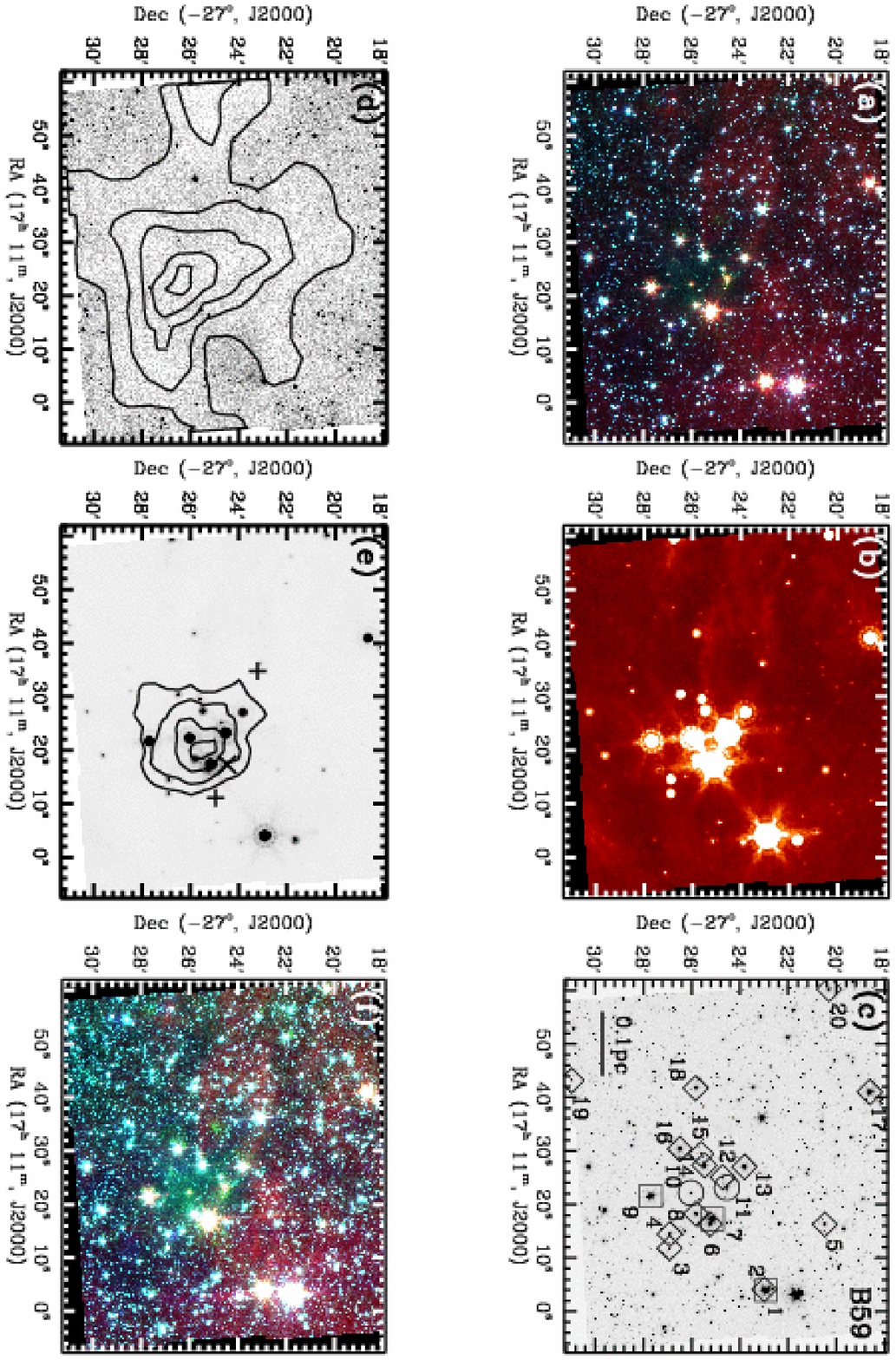}
\caption{}~\label{f1}
\end{figure}

\noindent
Fig 1 -- Multiwavelength views of the B59 region. {\it a)} 3-color
composite in IRAC filters: 3.6 $\mu$m ({\it blue}), 4.5 $\mu$m ({\it
green}), and 8.0 $\mu$m ({\it red}). {\it b)} MIPS 24 $\mu$m image
revealing the young star cluster.  Nonstar-like features seen $\pm2'$
roughly N and S of bright objects are artifacts due to latent
images. {\it c)} Source ID key for candidate young stars from Table 1,
with IRAC 4.5 $\mu$m image in background.  Young star IR spectral
classes have the following symbols: Class I ({\it circle}), Flat ({\it
square}), and Class II ({\it diamond}). Scale bar is for adopted
distance of $d=130$ pc.  {\it d)} Estimated A$_V$ from $2MASS$ data
with Digital Sky Survey Red image.  Extinction map has resolution
$\sim100$\arcsec ~and contours at A$_V$ = 10, 14, 19, 30, and 45
mag. {\it e)} MIPS 24 $\mu$m image with CS (J=2-1) integrated
intensity contours with a 60\arcsec ~beam from C. DeVries
(priv. comm.).  Contours are at 0.36, 0.50, 0.70, and 0.85 K
kms$^{-1}$.  CO red outflow ({\it dark x}), and CO blue outflow peaks
({\it dark crosses}) are from Onishi et al. (1999).  These are the
peak positions of the integrated line flux in red and blue channels.  A
third blue CO outflow peak is off the map to the SE. {\it f)} Same as
{\it (a)} but stretched to show extended emission.

\begin{figure}[!t]
\includegraphics[angle=0, scale=1.0]{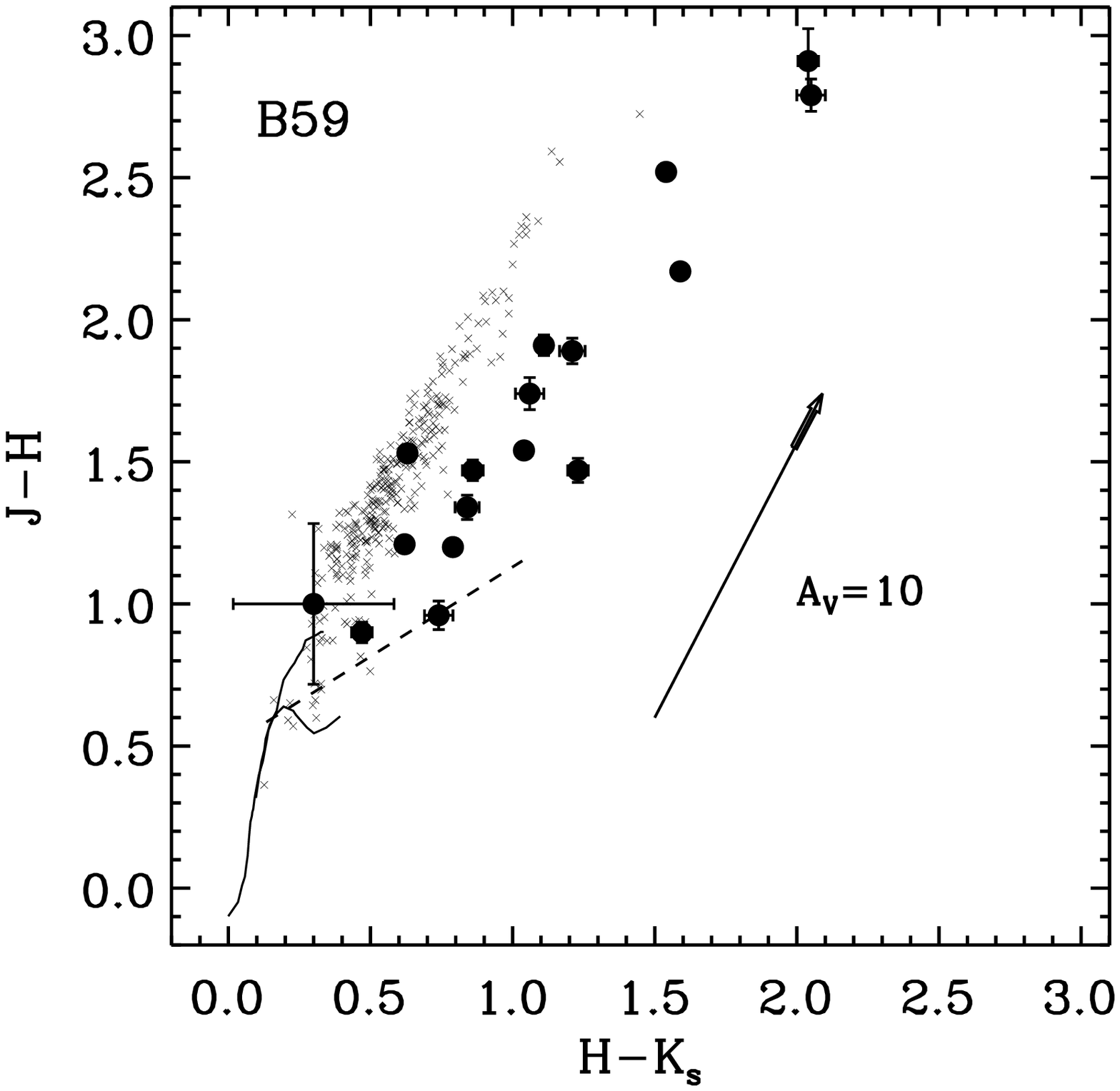}
\caption{2MASS JHK$_s$ fluxes of candidate young stars in B59 ({\it
filled circles}) and objects with J flux $\gtrsim2.5$ mJy and
classified as stars by {\it c2d} ({\it small x's}).  Main sequence and
giants from Bessell \& Brett (1988) are the {\it solid lines}, and the
Taurus T Tauri main locus from Meyer et al. (1997) is the {\it dashed
line}.  Both are converted to the 2MASS system using Carpenter (2000).
The reddening arrow shows the adopted extinction law (Indebetouw et
al., 2005) converted to A$_V$ (see text) and with A$_V$ = 10 mags.}
~\label{f2}
\end{figure}

\begin{figure}[!t]
\includegraphics[angle=0,
scale=0.9]{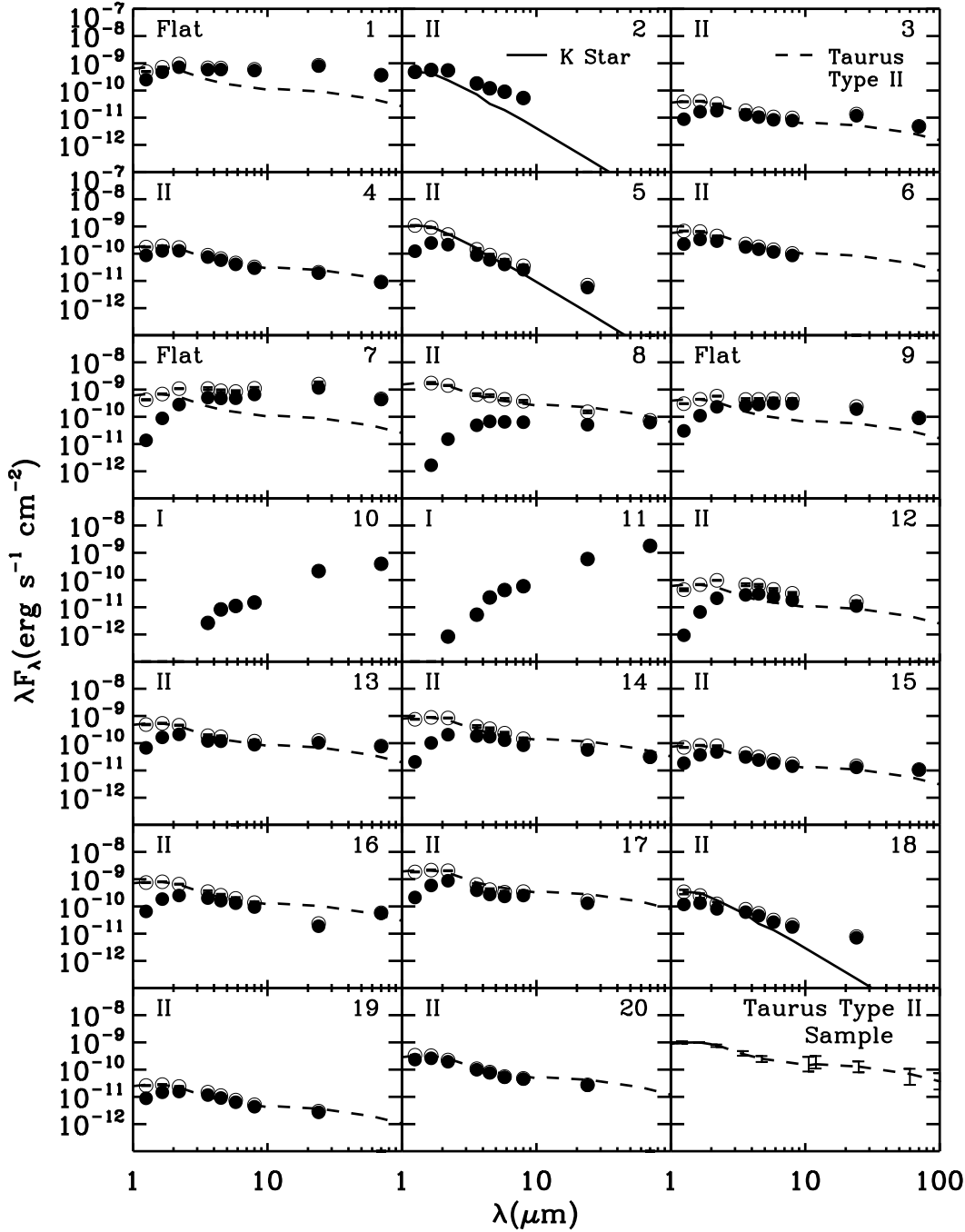}
\caption{Spectral energy distributions (SEDs) of the candidate young
stars in B59 as observed ({\it filled circles}) and after correction
for extinction as described in text ({\it open circles}).  Error bars,
including absolute calibration uncertainties, are shown, but are
smaller than the symbols.  In each object panel, the source number
(Table 1) is in the upper right and the IR spectral class after
correction for extinction, where possible, is in the upper left.  The
two Class I sources are not corrected for extinction.  For the Flat
and Class II sources with $\alpha \geq -1.6$, the median Class II SED
of Taurus sources ({\it lower right box}, D'Alessio et al. 1999)
scaled at H (1.65 $\mu$m), is plotted for comparison ({\it short
dashed lines}).  Class II sources with $\alpha < -1.6$ have a K7
photosphere scaled at J(1.25 $\mu$m) ({\it solid lines}).}~\label{f3}
\end{figure}

\begin{figure}[!t]
\includegraphics[angle=0, scale=1.0]{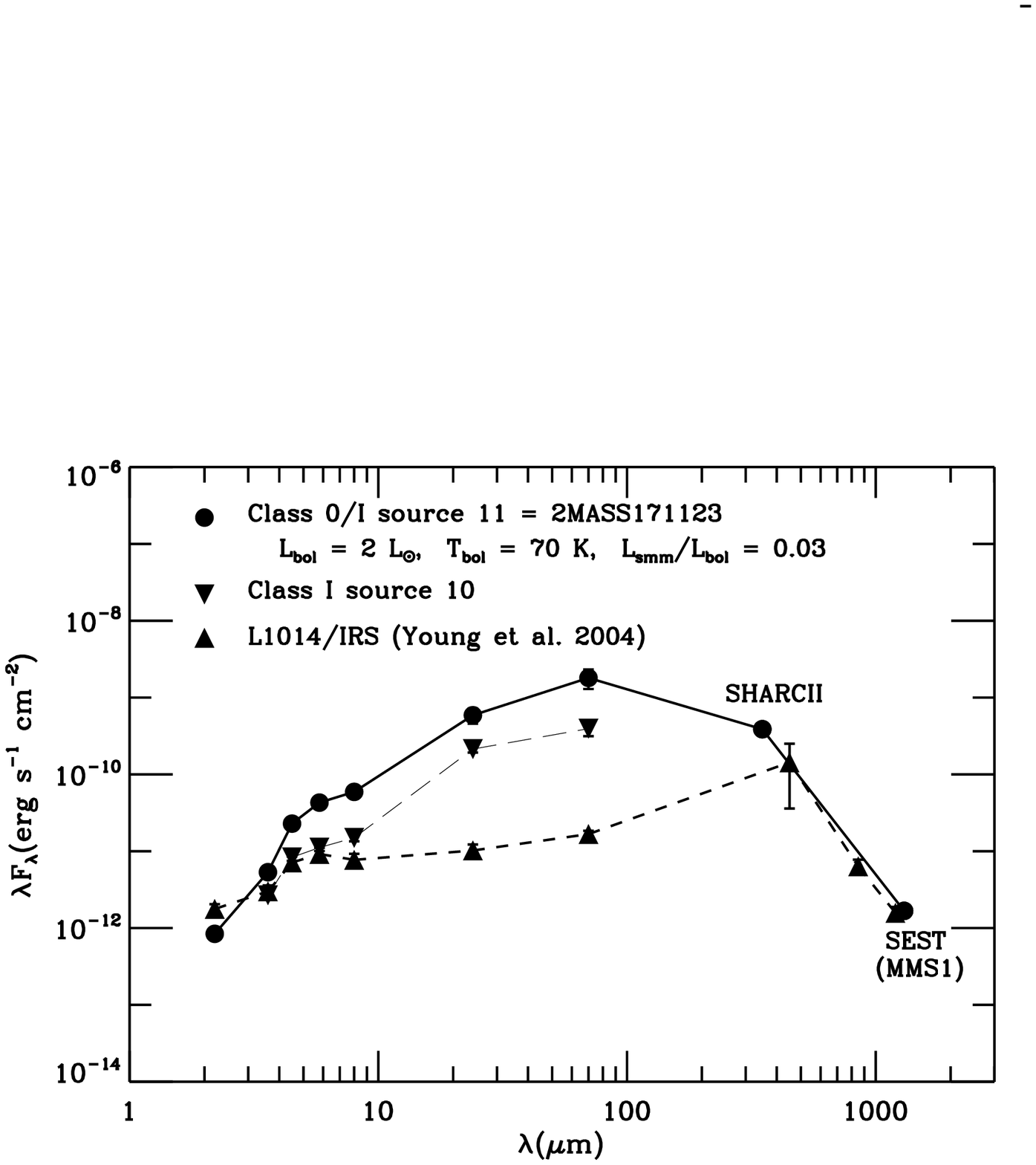}
\caption{Spectral energy distributions of the candidate Class 0/I or I
protostars in B59, compared to the infrared source in L1014.  For
2MASS171123, the SHARCII 350 $\mu$m data point is from Wu et al.
(2006, submitted to ApJ).  SEST 1.3mm data point of MMS1 is from
Reipurth et al. (1996).  See text for details. }~\label{f4}
\end{figure}

\begin{figure}[!t]
\centering
\includegraphics[angle=0, scale=1.0]{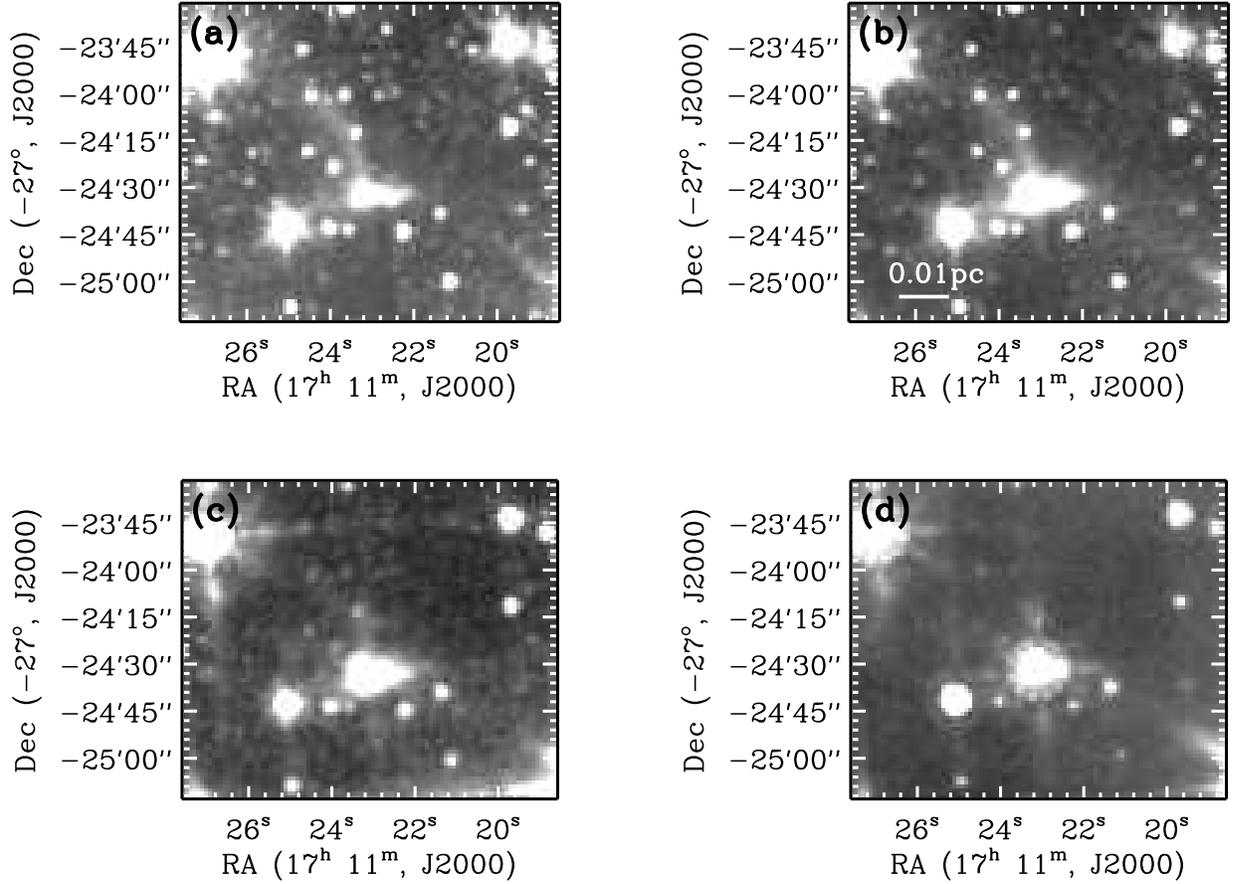}
\caption{Close-up of the Class 0/I source 11 = 2MASS171123, closest to
center, through the 4 IRAC bands: {\it a)} 3.6 $\mu$m, {\it b)} 4.5
$\mu$m, {\it c)} 5.8 $\mu$m, {\it d)} 8.0 $\mu$m.  Image arcs visible
at 3.6 and 4.5 $\mu$m may trace an outflow cavity in scattered light.
Scale bar plotted in {\it (b)} is for assumed distance of $d=130$
pc.}~\label{f5}
\end{figure}

\begin{figure}[!t]
\includegraphics[angle=0, scale=1.0]{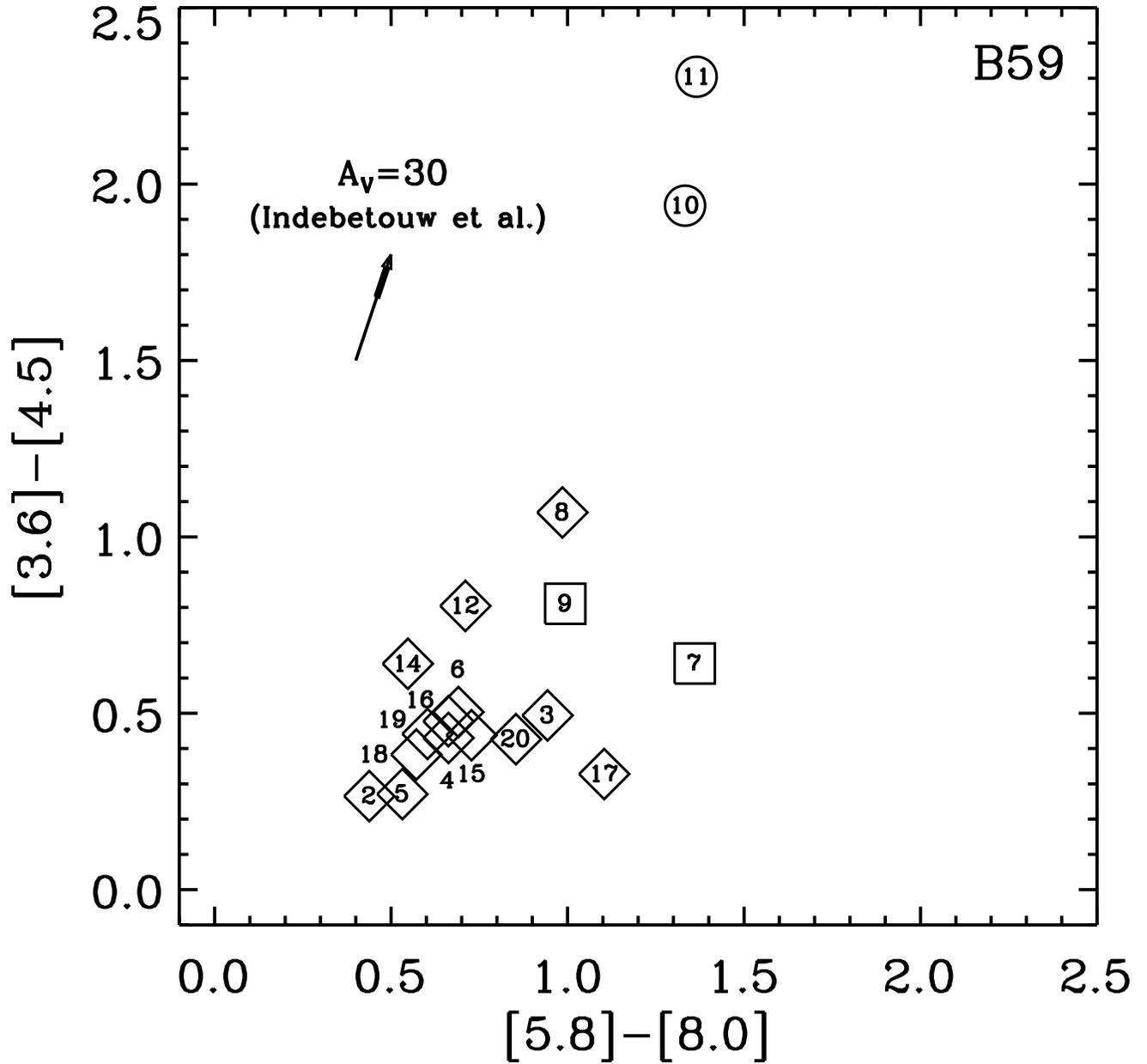}
\caption{IRAC color-color diagram, as observed, of candidate young
stars in B59 with source numbers from Table 1.  Young star IR spectral
classes, after correction for extinction (where possible) are
indicated with the following symbols: Class I ({\it circle}), Flat
({\it square}), and Class II ({\it diamond}).  The reddening arrow
shows the adopted extinction law (Indebetouw et al., 2005) for A$_V$ $\approx$
30 mags (see text).}~\label{f6}
\end{figure}

\clearpage

\begin{deluxetable}{rccllrrrrrl}
\tabletypesize{\scriptsize}
\rotate
\tablecaption{Candidate Young Stars in B59$^a$\label{tbl-1}}
\tablewidth{0pt}
\tablehead{
\colhead{No.} & \colhead{2MASS Name} & 
\colhead{Other ID} & \colhead{RA(J2000)} & \colhead{DEC(J2000)} &
\colhead{J} & \colhead{H} & \colhead{K$_s$} & \colhead{A$_V$(JHK)$^g$} &
\colhead{A$_V$(Map)$^h$} & \colhead{Comments}
}
\startdata
1 & J17110392-2722551 & LkH$\alpha$ 346 NW & 17\phn11\phn03.91 & 
  -27\phn22\phn55.2 & 10.46 & 8.99 & 7.76 & 2.6 & 10.0 & Triple$^b$ \\
  &                  &                    &                      &
                  & (0.03) & (0.03) & (0.02) & (1.7) & (0.6) & IRAS17079-2719 \\
2 & J17110411-2722593 & LkH$\alpha$ 346 SE & 17\phn11\phn04.12 & 
  -27\phn22\phn59.3 & 9.75 & 8.79 & 8.05 & 0.0 & 9.9 &
\\
  &                  &                    &                      &
                  & (0.03) & (0.04) & (0.03) & (1.5) & (0.6) & \\
\\
3 & J17111182-2726547 & \nodata    & 17\phn11\phn11.82 & 
  -27\phn26\phn55.0 & 14.09 & 12.62& 11.76 & 5.6 & 19.2 & \\
  &                  &                    &                      &
                  & (0.02)& (0.03)&(0.02)& (1.4) & (1.9) & \\
\\
4 & J17111445-2726543 & \nodata       & 17\phn11\phn14.45 & 
  -27\phn26\phn54.4 & 11.62& 10.42& 9.63 & 2.7 & 22.4 &
\\
  &                  &                    &                      &
                  & (0.02)& (0.02)&(0.02)& (1.3) & (2.8) & \\
\\
5 & J17111626-2720287 & \nodata    & 17\phn11\phn16.27 & 
  -27\phn20\phn28.8 & 11.23& 9.70 & 9.07 & 8.3 & 7.3 &
\\
  &                  &                    &                      &
                  & (0.02)& (0.02)&(0.02)& (0.8) & (0.3) & \\
\\
6 & J17111631-2725144 & KW 2               & 17\phn11\phn16.32 &
  -27\phn25\phn14.6 & 10.58 & 9.37 & 8.75 & 4.2 & 14.7 & H$\alpha^c$ \\
  &                  &                    &                      &
                    & (0.02)&(0.02) & (0.02) & (1.1) & (2.3) & \\
\\
7 & J17111726-2725081 & IRAS17081-2721     & 17\phn11\phn17.28 & 
  -27\phn25\phn08.2 & 13.61 & 10.82 & 8.77 & 13.0 & 15.7 & \\ 
  &                  &                    &                      &
                    & (0.04)&(0.04) & (0.03) & (1.8) & (2.5) & \\
\\
8 & J17111827-2725491 & \nodata    & 17\phn11\phn18.13 & 
  -27\phn25\phn49.3 & $>$18.4 & 15.12 & 11.95 & 44 & 25.3 & \\ 
  &                  &                    &                      &
                    &       &(0.08) & (0.03) &  (8) & (6.2) & \\
\\
9 & J17112153-2727417 & IRAS17082-2724 & 17\phn11\phn21.50 &
   -27\phn27\phn42.3 & 12.74 & 10.57 & 8.98  & 8.7 & 26.9 &
\\
  &                  &                    &                      &
                    & (0.02)&(0.02) & (0.02) & (1.7) & (5.0) & \\
\\
10 & \nodata & \nodata & 17\phn11\phn22.10 &
   -27\phn26\phn\phn2.0 & \nodata & \nodata&\nodata & \nodata & 46.1 &
\\
  &                  &                    &                      &
                    &       &       &        &       & (20.0) & \\
\\
11 & J17112317-2724315 & \nodata & 17\phn11\phn23.18 & 
  -27\phn24\phn31.5 & $>$18.8&$>$17.8&15.08 & \nodata & 25.7 &
Prob. MMS1$^d$ \\
  &                  &                    &                      &
                    &       &        & (0.14) &       & (5.7) & \\
\\
12 & J17112508-2724425 & \nodata & 17\phn11\phn25.08 & 
  -27\phn24\phn42.7 & 16.52 & 13.61 & 11.57 & 14.7 & 27.2 & \\
  &                  &                    &                      &
                    & (0.11)&(0.03) & (0.02) &  (2.2) & (5.7) & \\
\\
13 & J17112701-2723485 &   B59-1    & 17\phn11\phn26.95 & 
  -27\phn23\phn48.4 & 11.88 & 10.14 &  9.08 & 7.5 & 24.4 &
Triple$^e$ \\
  &                  &                    &                      &
                    & (0.04)&(0.04) & (0.03)&  (1.5) & (2.4) &
\\
\\
14 & J17112729-2725283 & \nodata & 17\phn11\phn27.06 & 
   -27\phn25\phn29.5 & 13.17 & 10.65 & 9.11 & 13.7 & 33.2 & \\
  &                  &                    &                      &
                    & (0.02)&(0.02) & (0.02)&  (1.3) & (8.6) & \\
\\
15 & J17112942-2725367 & \nodata & 17\phn11\phn29.31 & 
  -27\phn25\phn36.3 & 13.28 & 11.74 & 10.70 & 5.1 & 23.6 & \\
  &                  &                    &                      &
                    & (0.02)&(0.02) & (0.02)& (1.3) & (3.7) &  \\
\\
16 & J17113036-2726292 & \nodata      & 17\phn11\phn30.29 & 
  -27\phn26\phn29.3 & 11.91 & 10.00 & 8.89 & 9.3 & 20.7 &
\\
  &                  &                    &                      &
                    & (0.03)&(0.02) & (0.02)& (1.4) & (2.7) &  \\
\\
17 & J17114099-2718368 & IRAS17085-2715  & 17\phn11\phn40.99 & 
  -27\phn18\phn37.0 & 10.64 & 8.75 & 7.54 & 8.2 & 8.4 & \\
  &                  &                    &                      &
                    & (0.02)&(0.04) & (0.02)&  (1.4) & (0.8) & \\
\\
18 & J17114182-2725477 &  B59-2 & 17\phn11\phn41.73 & 
  -27\phn25\phn50.3 & 11.3 & 10.4 & 10.1 & 4.1 & 11.7 &
Double$^f$ \\
  &                  &                    &                      &
                    & (0.2)&(0.2) & (0.2)&  (4.3) & (0.6) & \\
\\
19 & J17114315-2730584 & \nodata   & 17\phn11\phn43.16 & 
  -27\phn30\phn58.6 & 14.09 & 12.75 & 11.91 & 4.1 & 10.1 & \\
  &                  &                    &                      &
                    & (0.03)&(0.03) & (0.03)& (1.4) & (1.2) & \\
\\
20 & J17120020-2720180 & LkH$\alpha$ 347   & 17\phn12\phn0.20 & 
  -27\phn20\phn18.1 & 10.54 & 9.64 & 9.17 & 1.4 & 6.9 & \\
  &                  &                    &                      &
                    & (0.02)&(0.03) & (0.02)& (1.1) & (0.8) & \\
\\

\enddata

\tablecomments{}

\tablenotetext{a}{ ~Within field covered by both IRAC and MIPS 24 $\mu$m.  
Uncertainties in parentheses.  Adopted flux zero points for the 2MASS J, H, 
and K$_s$ filters are 1594, 1024, and 667 Jy, respectively.}
\tablenotetext{b}{ ~LkH$\alpha$ 346 SE is labelled the primary (A) and 
has a tertiary companion (C) according to Chelli et al. (1995).  The 
AC pair is unresolved by {\it Spitzer}.  LkH$\alpha$ 346 NW, or (B), 
is the brighter infrared source.}
\tablenotetext{c}{ ~H$\alpha$ emission (Kohoutek \& Wehmeyer 2003).}
\tablenotetext{d}{ ~Probably MMS1 of Reipurth et al. (1996). See text.}
\tablenotetext{e}{ ~The source is triple according to Koresko (2002).
The brighter source is the primary (plus tertiary), for which the 
position and fluxes are given.}
\tablenotetext{f}{ ~The source is double (Reipurth \& Zinnecker 1993),
possibly triple (Koresko 2002).  The brighter mid-infrared source is the 
secondary, for which the position and fluxes are given.  The JHK 
fluxes are from Koresko (2002).}
\tablenotetext{g}{~Extinction from dereddening to T Tauri JHK$_s$ locus.}
\tablenotetext{h}{~Cloud extinction from background stars.}

\end{deluxetable}

\clearpage

\begin{deluxetable}{rcrrrrrrccccc}
\tabletypesize{\scriptsize}
\rotate
\tablecaption{Spitzer Fluxes (mJy) of Candidate Young Stars in 
B59$^a$\label{tbl-2}}
\tablewidth{0pt}
\tablehead{
\colhead{No.} & \colhead{Other ID} & 
\colhead{3.6} & \colhead{4.5} & \colhead{5.8} & \colhead{8.0} & \colhead{24} &
\colhead{70} & \colhead{$\alpha^b$} & \colhead{IR Class$^c$} & 
\colhead{$\alpha^d$} & \colhead{IR Class$^e$} & 
\colhead{L$_{\rm bol}$(L$_{\odot}$)$^f$}  
}
\startdata
1 & LkH$\alpha$ 346 NW & 699 & 881 & \nodata & 1440 & 6320 & 8490& -0.22 & Flat & -0.35$^g$ & Flat & $>$1.6 \\
  &                    & (9) & (13)&         & ( 25)&(1260)& (40)&&&&& \\
\\
2 & LkH$\alpha$ 346 SE & 218 & 180 &    174  &  141 &\nodata&\nodata& -1.81& II & -1.81 & II & 0.50 \\
  &                    &  (3)&  (2)&     (4) &   (1)&      &     &&&&&(0.11) \\
\\
3 & \nodata            & 15.5& 15.8& 16.2&20.9 & 94.5 & 110 & -0.64 & II & -0.92 & II & 0.14 \\
  &                    & (0.2)&(0.1)&(0.1)&(0.1)&(0.4)& (5) &     &&&&(0.08) \\
\\
4 & \nodata            & 90.1& 86.6& 78.4&78.1 & 155 & 210   & -1.14 & II & -1.27 & II & 0.26 \\
  &                    & (1.1)&(0.7)&(0.5)&(0.4)&(1)  &(6 )  &     &&&&(0.09) \\
\\
5 & \nodata            & 106& 88.0& 77.1& 68.1& 45.5 &  $<$100 & -1.64 & II & -2.04 & II & 0.98 \\
  &                    & (1)&(1)&(0.5)&(0.6)  &(0.3) &         &&&&&(0.09) \\
\\
6 & KW 2               & 211 & 217 & 223 & 228 & \nodata & \nodata & -0.94 & II & -1.14 & II & 0.93 \\
  &                    & (3) & (3) & (2) & (3) &         &       &&&&&(0.13) \\
\\
7 & IRAS17081-2721     & 609 & 711 &929 &1760 &   9200  &   10100 &  0.66 &  I & 0.02 & Flat & $>$2.7 \\
  &                    & (8)& (16)& (11)&(24) &  (1840) &    (50)&&&&& \\
\\
8 & \nodata            & 58.9 & 102&126  &169  &   412   & 1440 &  1.11 &  I & -1.04 & II & 1.4 \\
  &                    & (0.7)& (2)& (1) &(2)  &   (3)   & (10) &&&&&(1.0) \\
\\
9 & IRAS17082-2724     & 310  &423 &601  &812  &1540    &2070     & 0.21  & Flat & -0.22 & Flat & $>$0.86 \\
  &                    &(4)   & (6)& (6) &(9) &(17)     &(10)    &&&&& \\
\\
10 & \nodata          & 3.2 &12.5 &21.6 &39.9  &1710 & 9190 & 2.16$^g$ & I & \nodata & \nodata & 0.63 \\
  &                   &(0.1)& (0.1)& (0.2)&(0.2) &(29)& (30)&&&&&(0.13)\\
\\
11& \nodata            & 6.4 & 34.3& 83.0& 158.0& 4715 & 42100 & 3.29  & 0/I & \nodata & \nodata & 2.2 \\
  &                    &(0.2)& (0.8)& (0.8)&(1.8) &(940)& (8400)   &&&&&(0.3) \\
\\
12& \nodata            & 34.2& 46.4& 46.2 & 48.1& 88.7 & \nodata & -0.13 & Flat & -0.85 & II & 0.16 \\
  &                    &(0.3)& (0.4)&(0.3)&(0.3) & (0.5) &          &&&&&(0.09) \\
\\
13& B59-1              & 146& 178 & \nodata & 235 & 831 & 1780 & -0.68 & II & -1.04 & II & 0.53 \\
  &                    &(2) & (2)&          & (3) & (5) &  (10)&&&&&(0.13) \\
\\
14& \nodata            & 222& 259 & 250 & 224 & 455 & 710 & -0.69 & II & -1.36 & II & 0.81 \\
  &                    & (3)& (3) & (3) & (3) & (3) & (7) &&&&&(0.16) \\
\\
15& \nodata            & 37.5 & 36.3 & 36.2 & 38.3 & 104 & 246 & -0.93 & II & -1.18 & II & 0.17 \\
  &                    & (0.3)& (0.3)& (0.3)& (0.3)&(1) &(4)&&&&&(0.09) \\
\\
16& \nodata            & 243  & 244  & 254 & 253  & 150 & 1310 & -0.76 & II & -1.21 & II & 0.74 \\
  &                    & (3)  & (3)  & (2) & (3)  & (1) &(6)&&&&&(0.16) \\
\\
17& IRAS17085-2715     & 471  & 412  & 450 & 670  & 1050 & \nodata & -0.97 & II & -1.37 & II & 1.81 \\
  &                    & (7)  & (6)  & (4) & (8)  & (9) &        &&&&&(0.37) \\
\\
18& B59-2              & 74.1  & 68.2  & 51.2    & 46.9  & 56.5 & $<$100 & -1.57$^h$ & II & -1.67$^h$ & II & 0.1-0.9 \\
  &                    & (0.8)  & (0.8)  & (0.3) & (0.4)  & (0.4) &   &&&&& \\
\\
19& \nodata            & 14.0  & 13.6  & 12.3    & 11.6  & 21.6 & $<$100 & -0.99 & II & -1.19 & II & 0.13 \\
  &                    & (0.1)  & (0.2)& (0.1)   & (0.1) & (0.2)&     &&&&&(0.08) \\
\\
20& LkH$\alpha$ 347    & 119  & 114  & 101   & 120  & 213 & \nodata & -1.13 & II & -1.20 & II & 0.36 \\
  &                    & (2)  & (2)  & (1)   & (1)  & (1) &     &&&&&(0.10) \\
\\

\enddata

\tablecomments{}

\tablenotetext{a}{ ~Within field covered by both IRAC and MIPS 24 $\mu$m.  1$\sigma$ errors
(in parentheses) are from photometric fitting and do not include absoulute flux error (see text).
A blank entry indicates no useful value due to confusion with nearby source(s), or due to
lack of data.}
\tablenotetext{b}{ ~Spectral index $d{\rm log}(\lambda F_{\lambda})/d{\rm log}
(\lambda)$ with endpoints at 2.2 and 8.0 $\mu$m uncorrected for extinction, 
unless different endpoint wavelengths in $\mu$m are noted.}
\tablenotetext{c}{ ~IR spectral class. }
\tablenotetext{d}{ ~Same as b after correction for extinction. }
\tablenotetext{e}{ ~Same as c after correction for extinction. }
\tablenotetext{f}{ ~Bolometric luminosity (or lower limit) estimated 
as discussed in text.  Uncertainties (in parentheses) do not include 
uncertainty in distance.}
\tablenotetext{g}{ ~On border of Flat and Class II, but longer wavelengths indicate Flat.}
\tablenotetext{h}{ ~Spectral index is from 3.6 to 8.0 $\mu$m. }

\end{deluxetable}

\end{document}